\documentclass[conference]{IEEEtran}
\IEEEoverridecommandlockouts
\usepackage{amsmath,epsfig}
\usepackage{algorithm}
\usepackage{bm}
\usepackage{tikz}
\usepackage{graphicx}
\usetikzlibrary{positioning}
\usetikzlibrary{shapes.multipart}
\usepackage{pgfplots}
\usepackage[usenames,dvipsnames]{pstricks}
\usepackage{epsfig}
\usepackage{pst-grad} 
\usepackage{pst-plot} 
\usepackage{subfigure}
\usepackage[capitalize]{cleveref}

\usepackage{subfigure}
\usepackage[T1]{fontenc}
\usepackage{amsmath}
\usepackage{graphics}
\include{psfig}
\usepackage{epsfig}
\usepackage{array}
\usepackage{psfrag}
\usepackage{amssymb}
\usepackage{color}
\usepackage{pstricks,pst-node,pst-text,pst-3d,pst-plot}
\usepackage{cite}

\usepackage{ulem}
\definecolor{dgreen}{rgb}{0, 0.8, 0.1}

\def\bA{{\bf A}}
\def\ba{{\bf a}}

\def\bs{{\bf s}}
\def\bS{{\bf S}}

\def\BibTeX{{\rm B\kern-.05em{\sc i\kern-.025em b}\kern-.08em
    T\kern-.1667em\lower.7ex\hbox{E}\kern-.125emX}}
\begin{document}

\title{Millimeter-Wave Circular Synthetic Aperture Radar Imaging}
\author{\IEEEauthorblockN{Shahrokh Hamidi}
\IEEEauthorblockA{Department of Electrical and Computer Engineering\\
University of Waterloo\\
Waterloo, Ontario, Canada\\
Email: shahrokh.hamidi@uwaterloo.ca}
\and
\IEEEauthorblockN{Safieddin Safavi-Naeini}
\IEEEauthorblockA{Department of Electrical and Computer Engineering\\
University of Waterloo\\
Waterloo, Ontario, Canada\\
Email: safavi@uwaterloo.ca}
}

\maketitle

\begin{tikzpicture}[remember picture, overlay]
      \node[font=\small] at ([yshift=-1cm]current page.north)  {This paper has been accepted for publication in the 2021 IEEE Canadian Conference on Electrical and Computer Engineering (CCECE). \copyright IEEE};
\end{tikzpicture}

\begin{abstract}
In this paper, we present a high resolution microwave imaging technique using a compact and low cost single channel Frequency Modulated Continuous Wave (FMCW) radar based on Circular Synthetic Aperture Radar (CSAR) technique. We develop an algorithm to reconstruct the image from the raw data and analyze different aspects of the system analytically.
Furthermore, we discuss the differences between the proposed systems in the literature and the one presented in this work.

Finally, we apply the proposed approach to the experimental data collected from a single channel FMCW radar operating at $\rm 79 \;GHz$ and present the results.
\end{abstract}


\section{Introduction}
Synthetic Aperture Radar (SAR) imaging is a well-known technique to produce high resolution radar images \cite{Cumming, Soumekh, shahrokh_sar1, shahrokh_sar2}. In the range direction the idea of pulse compression is used to achieve high resolution. In the azimuth direction, it is the relative motion of the radar with respect to the target that gives rise to higher resolution.
When the synthetic aperture is created circularly, the system is called Circular Synthetic Aperture Radar (CSAR). Previously, CSAR imaging has been addressed in the literature such as the airborne CSAR systems in \cite{Soumekh, Soumekh_2, CSAR_1, CSAR_2} in which the circular aperture is created around the scene to be imaged. In fact, the circular synthetic aperture encircles the scene. The Geostationary CSAR system operates in the same way \cite{CSAR_Geo}.
The ground based CSAR systems, a.k.a., ArcSAR, have also been studied extensively \cite{CSAR_G1, CSAR_G2, CSAR_G3, CSAR_G4, SAR_Truck}.
The ground based CSAR systems are different from the airborne and spaceborne CSAR systems in a way that the area to be imaged can be outside of the circular synthetic aperture which provides a unique opportunity to monitor the $360^o$ surroundings of the CSAR system using a small circular aperture.
Of course, we should mention that the Unmanned Aerial Vehicle (UAV) and Helicopter based CSAR systems, while stationary with respect to the ground, can also operate in the same mode as the ground based CSAR systems meaning that the scene to be imaged can be outside of the synthetic circular aperture exactly similar to the configuration we consider in this work.

In \cite{SAR_Truck}, the authors have addressed the previous work done in the field of CSAR imaging presented in \cite{CSAR_G1, CSAR_G2, CSAR_G3} and have developed a ground based CSAR system which is the same as the model we develop in this paper. They have created a circular aperture at X band. The scene to be imaged is outside of the synthetic aperture. Their CSAR system operates in two different modes which they have referred to as, scan mode and spot mode. The scan mode is referred to the case that the angle between the antenna axis and the circular track remains the same while the radar is collecting data from the scene to be imaged. In the spot mode, however, the antenna is spinning on its axis while it is moving along the track to direct the mainlobe on the target at all times during the data gathering process. For the image reconstruction they have used the Range Doppler algorithm.

In \cite{w_k_1, w_k_2, w_k_3}, sub-aperture processing, Range Doppler method as well as  $\omega-k$ algorithms have been used for the CSAR image reconstruction.
In \cite{CSAR_Hel}, a helicopter based CSAR imaging system has been developed to create high resolution images from the ground while the helicopter is stationary.

In this paper, we specifically focus on CSAR imaging in which the scene to be imaged is located outside of the aperture. We, further, consider the case in which the angle between the main axis of the antenna and the circular synthetic aperture remains fixed during the data gathering process. This allows us to cover the entire $360^o$ surroundings of the radar in one full rotation. As we mentioned before, our CSAR system is similar to that of \cite{SAR_Truck}. However, the algorithm that we develop to reconstruct the image is different. In \cite{SAR_Truck}, the Range Doppler algorithm has been used for image reconstruction. First, they perform range compression and then they have used the Taylor expansion of the radial distance between the radar and the target up to the second order to perform azimuth compression. The Range Doppler algorithm, however, can perform well only for narrow swath width in narrow beam mode. Furthermore, for radars with high range resolution, such as the one we consider in this work, the Range Cell Migration (RCM) phenomenon \cite{Cumming}, if not compensated for, can degrade the quality of the image considerably. The RCM effect refers to the case that the energy of a point target appears in different range cells per different azimuth lines. When we are dealing with small range cells (high range resolution) and gathering data with a wide beam over a large synthetic aperture, which is the case we consider in this paper, the RCM is unavoidable. In \cite{SAR_Truck}, the authors have not addressed the RCM phenomenon. We, on the other hand, develop a joint range-azimuth algorithm capable of forming a 2D image without considering any approximation. Our proposed algorithm is an exact solution for the CSAR imaging and it can operate for large swath width in wide beam mode. We, further, describe the angular sample spacing analytically which has not been addressed in \cite{SAR_Truck}.

In \cite{w_k_1, w_k_2, w_k_3}, the authors have used the sub-aperture processing, the Range Doppler, as well as the $\omega-k$ algorithms for CSAR image processing. Similar to the Range Doppler algorithm, the sub-aperture processing and the $\omega-K$ algorithm also suffer from the same issues we mentioned before.
To overcome the previously mentioned problems with the Range Doppler algorithm, the sub-aperture processing, and the $\omega-k$ method, in \cite{CSAR_Hel}, a technique based on the z chirp transform has been developed. However, the z chirp transform based technique proposed in \cite{CSAR_Hel}, is performed in 2D frequency domain, namely, the range frequency and the azimuth frequency domain which, as a result, the dependency of the RCM to the range is ignored \cite{Cumming}.

We should also mention that, despite the fact that we have conducted indoor experiments in this paper, the algorithm we have  developed in this work, can be used for outdoor ground based as well as UAV and Helicopter based CSAR imaging systems without any modification.

The structure of the paper is as follows. In section \ref{Model Description}, we describe the system model and formulate the problem. In section \ref{Image Reconstruction}, we present an algorithm for image reconstruction. In section \ref{Angular Sample Spacing}, we talk about the limits on the sample distancing in the angular domain.
Section \ref{Resolution Analysis} discusses the resolution limit.
Finally, section \ref{Experimental Results} has been dedicated to applying the proposed algorithm to the experimental data gathered from a single channel FMCW radar operating at $\rm 79 \;GHz$ followed by concluding remarks.

\section{Model Description}\label{Model Description}
\begin{figure}
\psfrag{BPM = 11}{\tiny {\rm{BPM = 11}}}
\centerline{
\includegraphics[height=5cm,width=6cm]{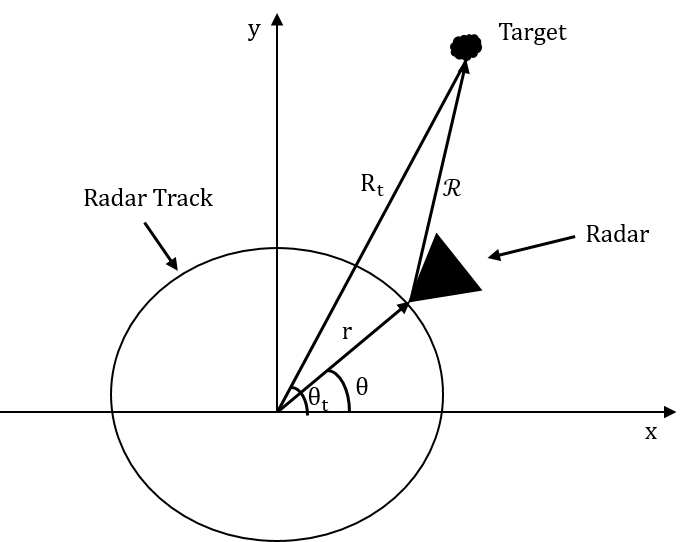}
\hspace{0.1cm}
}
\vspace*{0.1cm}
\caption{The geometry of the set-up. The radar track is a circle with radius $\rm r$.
\label{fig:Model_Geometry}}
\end{figure}

In this section, we develop the system model.
Fig.~\ref{fig:Model_Geometry} shows the  geometry of the model. The radar track is a circle with radius $\rm r$ on which the radar is moving with constant angular velocity. The radar transmits a Linear Frequency Modulated (LFM) signal, a.k.a., chirp signal, toward the target.
The reflected signal from the target is received at the location of the receiver and after being mixed with a copy of the transmitted signal, the result is expressed as
\begin{align}
\label{s}
\bs(\mathcal{R}) = [s(1,\mathcal{R}),s(2,\mathcal{R}),\cdots,s(N,\mathcal{R})]^T,
\end{align}
where $s(i)$ is given as
\begin{align}
\label{beat_signal}
s(i,\mathcal{R}) = \sigma(\mathcal{R}) e^{\displaystyle j2\pi[f_c - \frac{b}{2} + \frac{b}{N}(i-1)]\frac{2\mathcal{R}}{c}},
\end{align}
in which $b$ is the chirp bandwidth, $N$ is the number of time samples, $f_c$ and $c$ are the center frequency and the speed of light, respectively. Also, $i \in \{1,2,\cdots,N\}$ and $\sigma(\mathcal{R})$ is the complex valued reflection coefficient of a point reflector located at $\mathcal{R}$. The parameter $\mathcal{R}$ represents the radial distance between the target and the radar.

The transmitter and receiver of the FMCW radar are located at distance $\rm d$ from each other. Therefore, we are dealing with a Bistatic radar. However, if $d \ll \sqrt{4\alpha \frac{c}{f_c} \mathcal{R}}$, we can then use the Monostatic approximation in which $\mathcal{R}$ is measured with respect to the midpoint of the transmitter and receiver \cite{Bi_Mono}. The Monostatic approximation holds for the FMCW radar that we use in this paper. Therefore, our analysis will be based on Monostatic FMCW radar.

From Fig.~\ref{fig:Model_Geometry}, parameter $\mathcal{R}$ for a point reflector located at $(R_t,\tilde{\theta}_t)$ is described as
\begin{align}
\label{R}
\mathcal{R} = \sqrt{{R_t}^2 + {r}^2  - 2rR_t\cos(\theta - \tilde{\theta}_t)}.
\end{align}
Using (\ref{R}), we can rewrite (\ref{s}) for a point target located at $(R_t,\tilde{\theta}_t)$ as
\begin{align}
\label{signal}
\bs(k,\theta_m) = &\sigma(R_t,\tilde{\theta}_t) \times \nonumber \\
&e^{\displaystyle jk\sqrt{{R_t}^2 + {r}^2  - 2rR_t\cos(\theta_m - \tilde{\theta}_t)}},
\end{align}
where the $i^{\rm th}$ element of the vector $k$ is given as $k_i = \displaystyle 4\pi\frac{[f_c - \frac{b}{2} + \frac{b}{N}(i-1)]}{c} \; \rm for \; i \in \{1,2,\cdots,N\}$. The subscript index $m$ refers to the $m^{\rm th}$ angle over which the data has been collected.

\section{Image Reconstruction based on Time Domain Correlation Method} \label{Image Reconstruction}
To set the stage for image reconstruction process, we define the following $N \times M$ matrix
\begin{align}
\label{A}
& \bA(k,\hat{R_l},\hat{\theta}_l,\mathbb{N},\Theta) = \nonumber \\
&[\ba_1(k,\hat{R_l},\hat{\theta}_l), \ba_2(k,\hat{R_l},\hat{\theta}_l), \dots, \ba_M(k,\hat{R_l},\hat{\theta}_l)],
\end{align}
where $\ba_m(k,\hat{R_l},\hat{\theta}_l)$ is a $N \times 1$ vector which is given as $\ba_m(k,\hat{R_l},\hat{\theta}_l) = e^{\displaystyle jk\sqrt{\hat{R}^2_l + {r}^2  - 2r\hat{R_l}\cos(\theta_m - \hat{\theta}_l)}}$. Moreover, in (\ref{A}),  $\mathbb{N} = \{1,2,\cdots,N\}$ and $\Theta = \{\theta_1,\theta_2,\cdots,\theta_M\}$ is the range of angles over which the data is collected.

Furthermore, we define the following $N \times M$ matrix
\begin{align}
\label{S}
\bS(k, \Theta) = [\bs(k,{\theta}_1),\bs(k,{\theta}_2),\cdots,\bs(k,{\theta}_M)],
\end{align}
where $\bs(k,{\theta}_m)$ is given in (\ref{signal}).

\begin{figure}[htb]
\centering
\begin{tikzpicture}
  \node (img1){\includegraphics[scale=0.8]{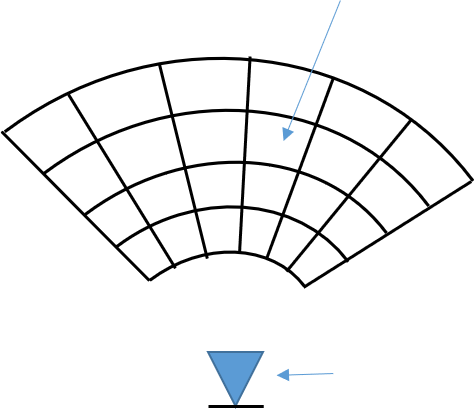}};
  \node[right=of img1, node distance=0cm, xshift=-3.1cm, yshift=-2.3cm,font=\color{black}] {\textit{Radar}};

\node[above=of img1, node distance=0cm, xshift=1.4cm, yshift=-1.2cm,font=\color{black}] {$(\hat{R},\hat{\theta})$};

\end{tikzpicture}
\caption{The figure shows the ROI which is composed of range of cells for hypothetical targets.}
\label{fig:ROI}
\end{figure}

The next step is to define the Region Of Interest (ROI) for the problem which has been shown in Fig.~\ref{fig:ROI}.
Finally, we express the image reconstruction algorithm as
\begin{align}
\label{CBP}
\mathcal{I}(\hat{R_l},\hat{\theta}_l) = & \mathbf{tr}[(\bS(k, \Theta) \bigodot \bA^{\dagger}(k,\hat{R_l},\hat{\theta}_l,\mathbb{N},\Theta))\nonumber \\
& *(\mathbf{1}_{M\times 1}*\mathbf{1}_{N\times 1}^T)],
\end{align}
in which $\mathcal{I}(\hat{R_l},\hat{\theta}_l)$ is the intensity for a hypothetical target located at $(\hat{R_l},\hat{\theta}_l)$ in the ROI. In (\ref{CBP}), $\mathbf{tr}$ stands for the trace of matrix, ${(.)}^\dagger$ and ${(.)}^T$ represent the complex conjugate and the transpose operators, respectively. Moreover, $\bigodot$ is the Hadamard matrix product (element-wise product), $*$ stands for the matrix product and $\mathbf{1}_{m\times 1}$ is a $m \times 1$ column vector with all its elements equal to 1. Also, the matrices $\bA(k,\hat{R_l},\hat{\theta}_l,\mathbb{N},\Theta)$ and $\bS(k, \Theta)$ are given in (\ref{A}) and (\ref{S}), respectively.
\section{Angular Sample Spacing}\label{Angular Sample Spacing}
In this section, we discuss the distance between samples taken in angular direction.

For $R_t \gg r$, we can use paraxial approximation and rewrite the phase term for the signal given in (\ref{signal}) as
\begin{align}
\label{phase}
\phi(\theta) = k[R_t + r\cos(\theta - \tilde{\theta}_t)].
\end{align}
Using $x = r\cos(\theta)$ and $y = r\sin(\theta)$ for the location of the radar, we can describe (\ref{phase}) as
\begin{align}
\label{phase_1}
\phi(x,y) = k[R_t + x\cos(\tilde{\theta}_t) + y\sin(\tilde{\theta}_t)].
\end{align}
Therefore, we can describe the wavenumber of the signal in $x$ direction as
\begin{align}
\label{Kx}
k_x & = \frac{\partial [k(R_t + x\cos(\tilde{\theta}_t) + y\sin(\tilde{\theta}_t))]}{\partial x},\nonumber \\
            & = k\cos(\tilde{\theta}_t).
\end{align}
As a result, the spatial distance between samples in $x$ direction is expressed as
\begin{align}
\label{delta_x}
\Delta_x & \leq \frac{2\pi}{2[(k_x)_{\rm max} - (k_x)_{\rm min}]}\nonumber \\
        & = \frac{c}{2\cos(\tilde{\theta}_t)[f_{\rm max} - f_{\rm min}]}.
\end{align}
Using $\mid {\Delta}_x \mid = r\sin(\theta)\Delta_{\theta}$, we can write (\ref{delta_x}) as
\begin{align}
\label{delta_theta}
\Delta_\theta  \leq  \frac{c}{r\sin(2\tilde{\theta}_t)[f_{\rm max} - f_{\rm min}]}.
\end{align}
From (\ref{delta_theta}), we infer that to satisfy the Nyquist rate in the $\theta$ direction the maximum sample spacing is given as
\begin{align}
\label{delta_theta_max}
\Delta_\theta  \leq  \frac{c}{r[f_{\rm max} - f_{\rm min}]}.
\end{align}
\section{Resolution Analysis}\label{Resolution Analysis}
In this section, we analyse the resolution of the system in both the range and angular domains.
The range resolution depends on the bandwidth of the transmitted signal and it is given as
\begin{align}
\label{resolution_r}
\delta_R  = \frac{c}{2b}.
\end{align}
Regarding the angular resolution, we know that for a linear SAR system the angular resolution is described as $\delta_a = \frac{\lambda}{2L_s}$ in which $L_s$ is the length of the synthetic aperture and $\lambda=\frac{c}{f_c}$ represents the wavelength of the signal \cite{Cumming, Soumekh}. In the case of CSAR imaging, the length of the synthetic aperture is $r\times \theta_{\rm 3dB}$ where $\theta_{\rm 3dB}$ is the $\rm 3\;dB$ beamwidth of the antenna. Therefore, the angular resolution for the CSAR system is expressed as
\begin{align}
\label{resolution_a}
\delta_a  = \frac{\lambda}{2r\times \theta_{\rm 3dB}}.
\end{align}
In fact, the angular resolution depends on the exposure time for a point target and from (\ref{resolution_a}), we can see that the angular resolution for CSAR system depends on $r$ and $\theta_{\rm 3dB}$ since a combination of both parameters will determine the exposure time.
\section{Experimental Results}\label{Experimental Results}
\begin{figure}[htb]
\centering
\begin{tikzpicture}
  \node (img1)  {\includegraphics[scale=0.7]{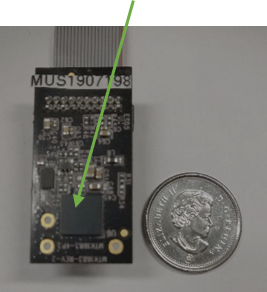}};
  \node[above=of img1, node distance=0cm, xshift=-0.4cm, yshift=-1.3cm,font=\color{black}] {{The radar chip}};
\end{tikzpicture}
\caption{The single channel FMCW radar from MediaTek company.}
\label{fig:Radar}
\end{figure}

\begin{figure}[htb]
\centering
\begin{tikzpicture}
  \node (img1)  {\includegraphics[scale=0.5]{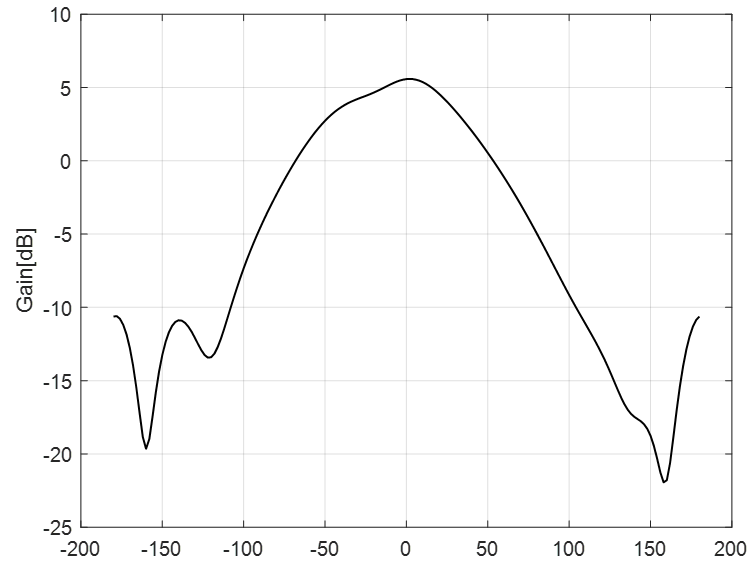}};
  \node[below=of img1, node distance=0cm, xshift=0cm, yshift=1.2cm,font=\color{black}] {$\theta^o$};
\end{tikzpicture}
\caption{The antenna pattern in the azimuth plane. The $\rm 3dB$ beamwidth and maximum gain at boresight are approximately $\rm 100^o$ and $\rm 5\;dB$, respectively.}
\label{fig:antenna_pattern}
\end{figure}

In this section, we apply the algorithm that we have developed to the experimental data.
Fig.~\ref{fig:Radar} shows the single channel FMCW radar from MediaTek company with on-chip antennas that we have used for our experiments.
Fig.~\ref{fig:antenna_pattern} illustrates the antenna pattern in azimuth plane.
As can be seen from Fig.~\ref{fig:antenna_pattern}, the $\rm 3 \;dB$ beamwidth is around $\rm 100^o$.
The center frequency of the radar is $\rm 79 \;GHz$ and the bandwidth has been set to $\rm 3.49 \; GHz$ which results in $\rm \delta_R = 4.3\; cm$ resolution in the range direction. The chirp time is $\rm 68.8 \; \mu s$. We have set the maximum range of the radar to $\rm 5.5 \; m$. We have $\rm 128$ samples per each chirp which means $\rm N=128$.

We have chosen $\rm r=13 \; cm$ for the radius of the circle over which the radar is rotating. As a result, for a target that is fully exposed by the $\rm 3\;dB$ beamwidth of the antenna we obtain $\delta_a = 0.24^o$ angular resolution. The choice of $\rm r=13 \; cm$ is not unique. However, it is important to be aware of the dependency of the angular resolution, the angular sample distancing and the size of the synthetic aperture on the parameter $r$ as we have explained them clearly in the paper.

The range of the angles over which we have taken the data, is $\rm {180}^o$.
Based on (\ref{delta_theta_max}), the angular step size should be $\Delta_\theta  \leq 37.82^o$. In our experiments, we have set the angular step size to $\rm {0.2}^o$ which as a result, the number of samples taken in angular direction is $\rm M=900$.

The data collection can be performed continuously or in stop-and-go mode. In the stop-and-go approach, a chirp  signal is transmitted toward the scene to be imaged and reflections are collected before the system moves to the next angular position. This process is repeated until the entire range of angles are covered.

In the continuous mode, the rate of chirp transmission is calculated based on the rotational speed of the robot in order for the system to be able to collect data at each angular step.

Regardless of which mode is used, during the data collection process, the received signals are stored in a matrix and when the data collection stage is finished, the data will then be  fed to the signal processing unit for image reconstruction.

\begin{figure}[htb]
\centering
\begin{tikzpicture}
  \node (img1)  {\includegraphics[scale=0.5]{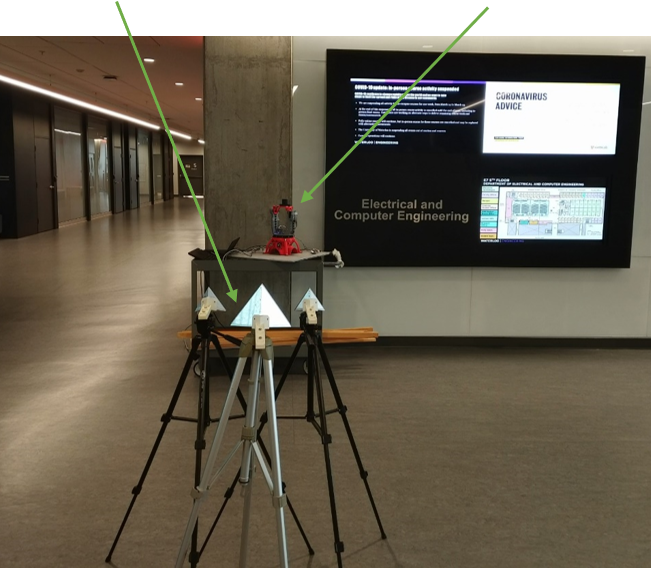}};
  \node[above=of img1, node distance=0cm, xshift=-2cm, yshift=-1.2cm,font=\color{black}] {{Corner reflectors}};
  \node[above=of img1, node distance=0cm, xshift=2cm, yshift=-1.2cm,font=\color{black}] {{The radar mounted on the robot}};
\end{tikzpicture}
\caption{The experimental set-up composed of the single channel FMCW radar mounted on the robot as well as a $\rm 20 \;dbsm$ and two 10dbsm corner reflectors.}
\label{fig:set-up}
\end{figure}

\begin{figure}
\centerline{
\includegraphics[height=3cm,width=3.5cm]{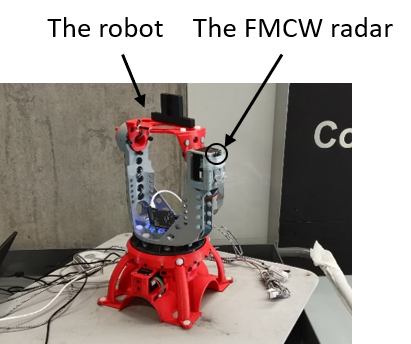}
\hspace{0.1cm}
\includegraphics[height=2cm,width=3.5cm]{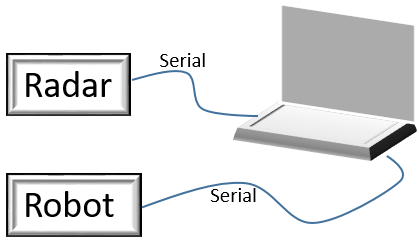}
\hspace{0.1cm}
}
\centerline{(a)\hspace*{4cm}(b)}
\vspace*{0.1cm}
\caption{a) the single channel FMCW radar mounted on the robot, b) A schematic which shows the connection between the radar, the robot, and the laptop.
\label{fig:radar_stepper}}
\end{figure}

The experimental set-up is shown in Fig.~\ref{fig:set-up}. A $\rm 20 \; dbsm$ along with two $\rm 10 \; dbsm$ corner reflectors have been chosen as targets.
To create the ROI shown in Fig.~\ref{fig:ROI}, we have chosen $0\leq \hat{R} \leq 4 \;m$ and $0^o\leq \hat{\theta} \leq 90^o$.
Fig.~\ref{fig:radar_stepper}-(a) shows the radar mounted on the robot in close-up.
The robot rotates the radar in a circular pattern. The distance from the center of the robot to the radar is $\rm r = 13 \;cm$. Fig.~\ref{fig:radar_stepper}-(b) presents a schematic which shows the connection between the radar, the robot, and the laptop.

The radar and the robot are connected to the laptop through serial port and they have been synchronised. The synchronization has been done in python. A single python code sends commands to both the radar and the robot. The received data from the radar is stored for later analysis which results in the final image formation.

\begin{figure}[htb]
\centering
\begin{tikzpicture}
  \node (img1)  {\includegraphics[scale=0.45]{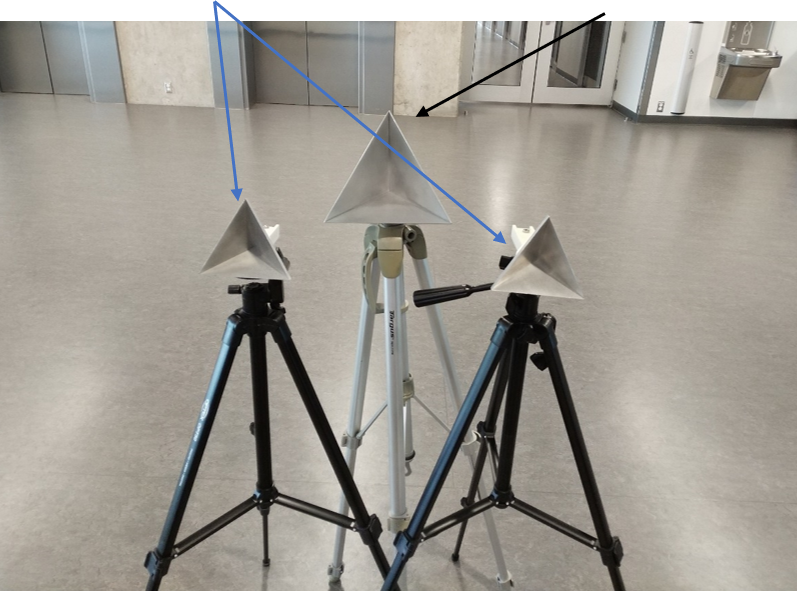}};
  \node[above=of img1, node distance=0cm, xshift=-2.4cm, yshift=-1.2cm,font=\color{black}] {{$\rm 10 \; dbsm$ corner reflectors}};
  \node[above=of img1, node distance=0cm, xshift=2.1cm, yshift=-1.2cm,font=\color{black}] {{The $\rm 20 \; dbsm$ corner reflector}};
\end{tikzpicture}
\caption{A $\rm 20 \; dbsm$ along with two $\rm 10 \;dbsm$ corner reflectors.}
\label{fig:targets}
\end{figure}

\begin{figure}
\psfrag{X} {\rm x}
\psfrag{y} {\rm y}
\centerline{
\includegraphics[height=5cm,width=6cm]{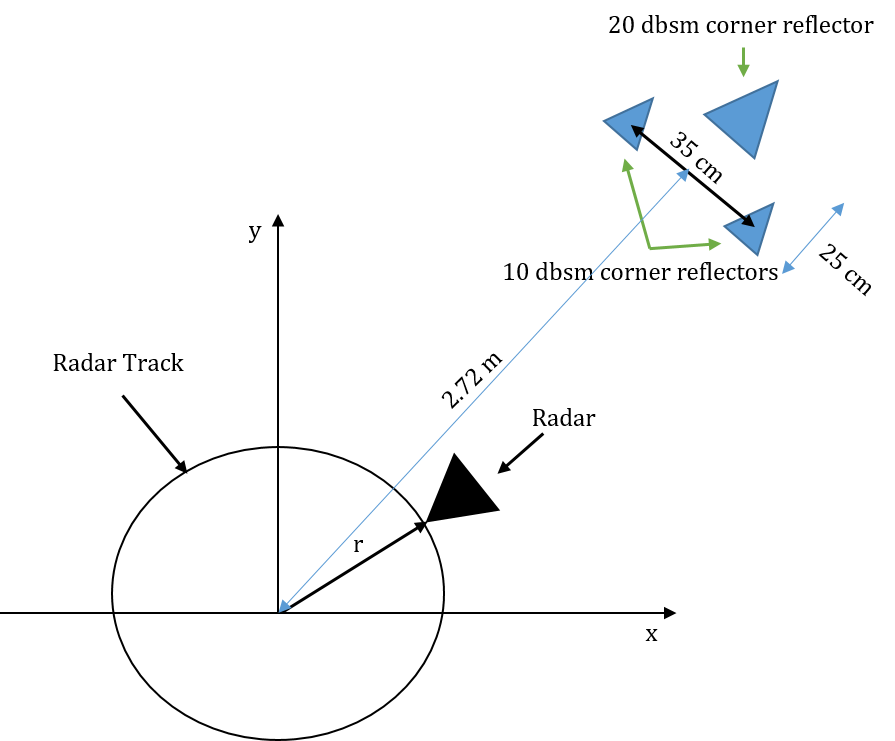}
\hspace{0.1cm}
}
\vspace*{0.1cm}
\caption{The position of a $\rm 20 \; dbsm$ as well as two $\rm 10\;dbsm$ corner reflectors in front of the radar used to collect the experimental data based on the set-up shown in Fig.~\ref{fig:set-up}.
\label{fig:targets_distance2}}
\end{figure}

\begin{figure}[htb]
\centering
\begin{tikzpicture}
  \node (img1)  {\includegraphics[scale=0.5]{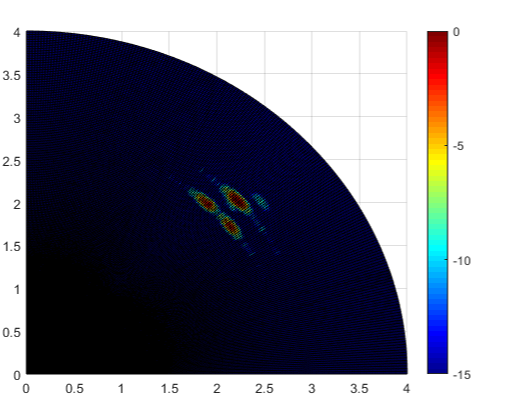}};
  \node[left=of img1, node distance=0cm, xshift=1.1cm, yshift=0cm,font=\color{black}] {y[m]};
  \node[below=of img1, node distance=0cm, xshift=-0.1cm, yshift=1cm,font=\color{black}] {x[m]};
\end{tikzpicture}
\caption{The reconstructed image based on the experimental data gathered by the set-up shown in Fig.~\ref{fig:set-up} and using the method given in (\ref{CBP}). The colorbar is in dB scale.}
\label{fig:Image}
\end{figure}

Fig.~\ref{fig:targets} illustrates the corner reflectors located in front of the radar.
Fig.~\ref{fig:targets_distance2} shows the corner reflectors and their distances from one another as well as from the center of the coordinate system.
Finally, the reconstructed image based on (\ref{CBP}) has been presented in Fig.~\ref{fig:Image}.

We have conducted another experiment using only two $\rm 10 \; dbsm$ corner reflectors. The horizontal distance between them is $\rm 24\;cm$. The radial distance from the center point of the two corner reflectors to the origin of the coordinate system is $\rm 4.73\;m$.

\begin{figure}
\psfrag{X} {\rm x}
\psfrag{y} {\rm y}
\centerline{
\includegraphics[height=5cm,width=5cm]{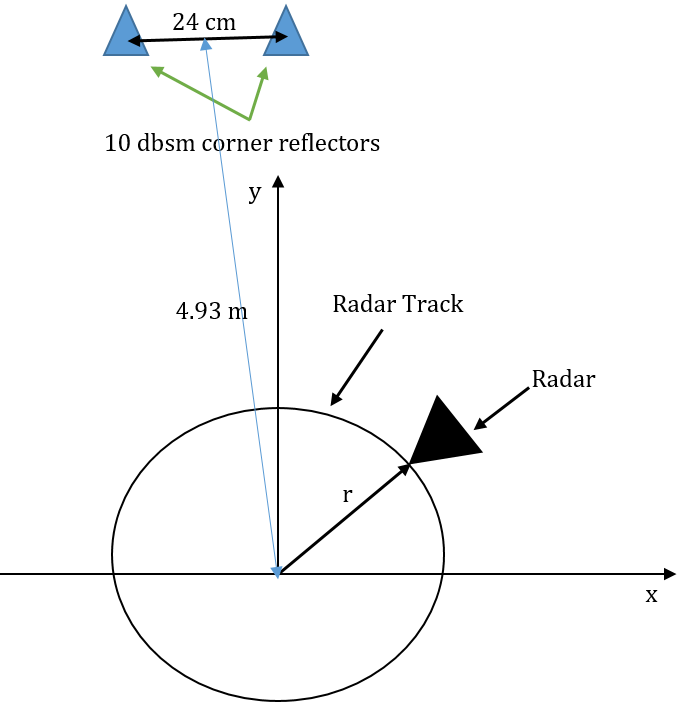}
\hspace{0.1cm}
}
\vspace*{0.1cm}
\caption{The position of the two $\rm 10 \; dbsm$ corner reflectors in front of the radar.
\label{fig:targets_distance3}}
\end{figure}

\begin{figure}[htb]
\centering
\begin{tikzpicture}
  \node (img1)  {\includegraphics[scale=0.35]{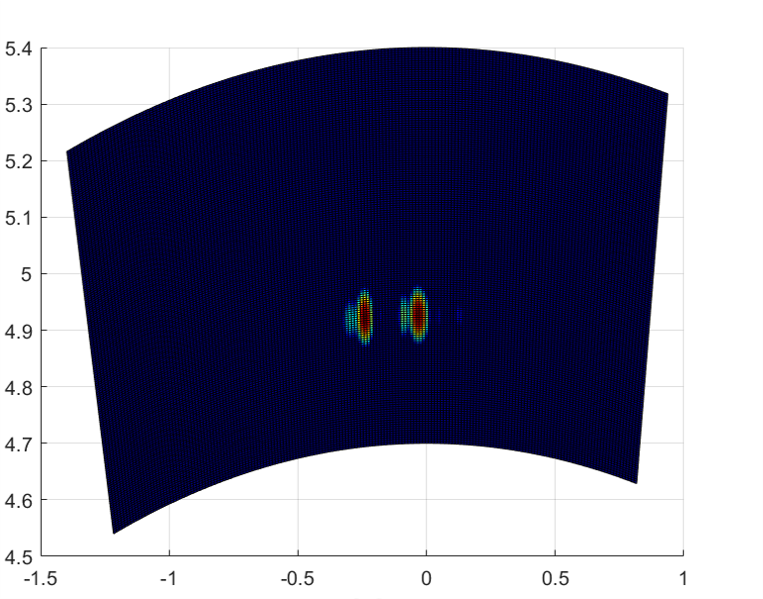}};
  \node[left=of img1, node distance=0cm, xshift=1cm, yshift=0cm,font=\color{black}] {y[m]};
  \node[below=of img1, node distance=0cm, xshift=0cm, yshift=1cm,font=\color{black}] {x[m]};
\end{tikzpicture}
\caption{The reconstructed image of two 10dbsm corner reflectors based on the experimental data gathered by the set-up shown in Fig.~\ref{fig:targets_distance3} and using the method given in (\ref{CBP}).}
\label{fig:Image_2}
\end{figure}

Fig.~\ref{fig:targets_distance3} illustrates the location of the corner reflectors with respect to each other and also with respect to the radar.
The reconstructed image for the experimental set-up described in Fig.~\ref{fig:targets_distance3} has been shown in Fig.~\ref{fig:Image_2}.
\section{Conclusions}
In this paper, we developed a high resolution imaging technique based on a compact and low cost single channel FMCW radar by utilizing circular motion. The imaging method allows $\rm 360^0$ coverage and the synthetic aperture is created over a small circle. We analysed the model at length and at the end applied the proposed technique to the experimental data gathered from a single channel FMCW radar operating at $\rm 79 \;GHz$ and discussed the results.

As a next step, we will be working on expanding the proposed algorithm to 3D imaging by utilising a linear array of antennas in the vertical direction.

Furthermore, the algorithm we developed in this paper, does not address image creation of moving targets. Therefore, another interesting and important topic to focus on, will be to address moving target imaging using CSAR.

\section{Acknowledgment}
The authors would like to thank MediaTek company for providing them with the FMCW radar.

\bibliographystyle{IEEEtran}
\bibliography{Biblio}

\begin{IEEEbiography}[{\includegraphics[width=1in,height=1.25in,clip,keepaspectratio]{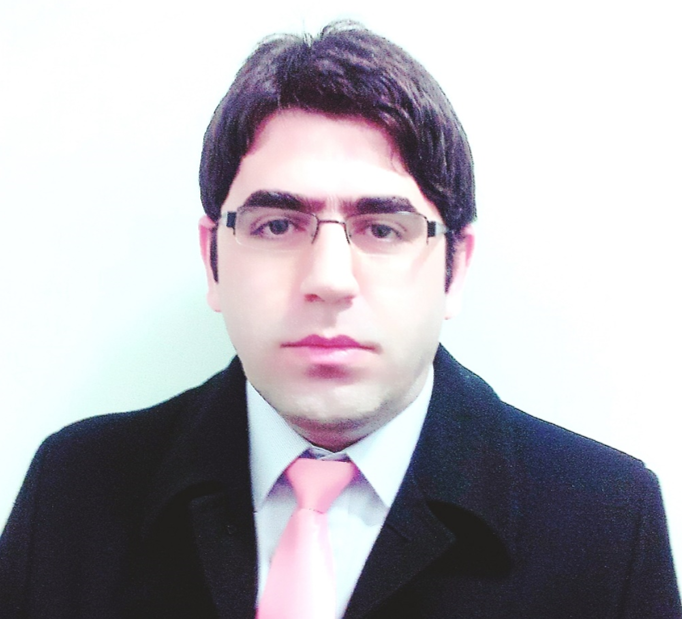}}]{Shahrokh Hamidi} was born in 1983, in Iran. He received his B.Sc., M.Sc., and Ph.D. degrees all in Electrical and Computer Engineering. He is with the faculty of Electrical and Computer Engineering at the University of Waterloo, Waterloo, Ontario, Canada. His current research areas include statistical signal processing, mmWave imaging, Terahertz imaging, image processing, system design,  multi-target tracking, wireless communication, machine learning, optimization, and array processing.
\end{IEEEbiography}

\begin{IEEEbiography}[{\includegraphics[width=2.5cm,height=3.5cm,clip,keepaspectratio]{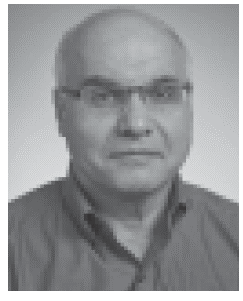}}]%
{SAFEDDIN SAFAVI-NAEINI}
 was born in Gachsaran, Iran, in 1951. He received the B.Sc. degree
in electrical engineering from the University of
Tehran, Tehran, Iran, in 1974, and the M.Sc.
and Ph.D. degrees in electrical engineering from
the University of Illinois, Urbana Champaign,
in 1975 and 1979, respectively. He joined the
Department of Electrical and Computer Engineering, University of Tehran, as an Assistant Professor, in 1980, where he became an Associate
Professor, in 1988. In 1996, he joined the Department of Electrical and
Computer Engineering, University of Waterloo, ON, Canada, where he is
currently a Full Professor and the RIM/NSERC Industrial Research Chair of
intelligent radio/antenna and photonics. He is also the Director of a newly
established Center for Intelligent Antenna and Radio System (CIARS).
He has published over 80 journal papers and 200 conference papers in
international conferences. His research activities deal with RF/microwave
technologies, smart integrated antennas and radio systems, mmW/THz
integrated technologies, nano-EM and photonics, EM in health sciences
and pharmaceutical engineering, antenna, wireless communications and
sensor systems and networks, new EM materials, bio-electro-magnetics,
and computational methods. He has led several international collaborative
research programs with research institutes in Germany, Finland, Japan,
China, Sweden, and USA.
\end{IEEEbiography}

\end{document}